\documentclass[amsmath,amssymb]{revtex4}
\usepackage{graphicx}
\usepackage{amscd,amsmath,amsthm,amsfonts,amssymb}
\def\PT{$\cal{PT}$}
\def\[{\begin{equation}}
\def\]{\end{equation}}

\begin{document}
\title{General rogue waves in the Boussinesq equation}
\author{Bo Yang and Jianke Yang}
\affiliation{Department of Mathematics and Statistics, University of Vermont, Burlington, VT 05405, USA}
\begin{abstract}
We derive general rogue wave solutions of arbitrary orders in the Boussinesq equation by the bilinear Kadomtsev-Petviashvili (KP) reduction method. These rogue solutions are given as Gram determinants with $2N-2$ free irreducible real parameters, where $N$ is the order of the rogue wave. Tuning these free parameters, rogue waves of various patterns are obtained, many of which have not been seen before. Compared to rogue waves in other integrable equations, a new feature of rogue waves in the Boussinesq equation is that the rogue wave of maximum amplitude at each order is generally asymmetric in space. On the technical aspect, our contribution to the  bilinear KP-reduction method for rogue waves is a new judicious choice of differential operators in the procedure, which drastically simplifies the dimension reduction calculation as well as the analytical expressions of rogue wave solutions.
\end{abstract}

\maketitle

\section{Introduction}
In 1871, Boussinesq introduced an equation which governs the propagation of long surface waves on water of constant depth \cite{JBoussinesq1871,JBoussinesq1872} (see also \cite{FUrsell1853}). After variable normalizations, this equation can be written as
\[ \label{Eqeta}
\eta_{tt}-\eta_{xx}-(\eta^2)_{xx}-\frac{1}{3}\eta_{xxxx}=0.
\]
Actually, the quantities $\eta, x$ and $t$ can be further scaled to produce any desired coefficients for the terms in this equation.
The present choice of coefficients is taken following \cite{ClarksonDowie2017}. This Boussinesq equation also arises in many other physical contexts, such as continuum approximations of certain Fermi-Pasta-Ulam chains~\cite{N.J.Zabusky1967,V.E.Zakharov1973,MtodaPR1975}, and ion sound waves in a plasma~\cite{A.C.Scott1975,E.Infeld1990}. Remarkably, this equation is integrable. Indeed, its multi-soliton solutions (by the bilinear method) and its Lax pair were reported almost simultaneously by Hirota and Zakharov in 1973 \cite{Hirota1973,V.E.Zakharov1973} (see also \cite{AblowitHaberm1975,DeiftTomei1982,AblowitClarks1991}).

Rogue waves are spontaneous large nonlinear waves which ``come from nowhere and disappear with no trace" \cite{Akhmediev2009}. In other words, they arise from a constant (uniform) background, run to high amplitudes, and then retreat back to the same background. These waves are associated with freak waves in the ocean and extreme events in optical fibers \cite{Ocean_rogue_review,ocean_book,Solli_optical,Wabnitz_book}.
Thus, they have attracted a lot of attention in the physics and nonlinear waves communities in recent years. Analytical expressions of rogue waves have been derived in a large number of integrable nonlinear wave equations, such as the nonlinear Schr\"odinger (NLS) equation \cite{Peregrine,AAS2009,DGKM2010,ACA2010,DPMB2011,KAAN2011,GLML2012,OhtaJY2012,DPMVB2013,HeNLSheight}, the derivative NLS equation \cite{XuHW2011,GLML2013}, the three-wave interaction equation \cite{BCDL2013}, the Davey-Stewartson equations \cite{OhtaJKY2012,OhtaJKY2013}, and many others \cite{AANJM2010,OhtaJKY2014,ASAN2010,TaoHe2012,BDCW2012,ManakovDark,Chow,PSLM2013,GmuZQin2014,Grimshaw_rogue,MuQin2016,LLMSasa2016,
LLMFZ2016,Degasperis1,Degasperis2,YangYang2018,XiaoeYong2018,Chen_Juntao,SunLian2018,YangYang2019}. Rogue waves have also been observed in water tanks \cite{Tank1,Tank2} and optical fibers \cite{Fiber1,Fiber2,Fiber3}.

In this paper, we consider rogue waves in the Boussinesq equation (\ref{Eqeta}). Since these waves arise from and retreat back to a constant background, we let
\[ \label{eta_bc}
\eta(x,t) \to \eta_0, \quad x, t \to \pm \infty,
\]
where $\eta_0$ is a constant. For rogue waves to appear, the background solution $\eta=\eta_0$ must be unstable to long-wave perturbations \cite{ManakovDark}. Simple calculations show that this so-called baseband instability occurs when $1+2\eta_0\le 0$. Under this condition, after a variable shift of $\eta=\eta_0+u$ and proper scalings of $x, t$ and $u$,
the Boussinesq equation (\ref{Eqeta}) reduces to
\[ \label{BoussiEq}
u_{tt}+u_{xx}-(u^2)_{xx}-\frac{1}{3}u_{xxxx}=0,
\]
and boundary conditions (\ref{eta_bc}) reduce to
\[ \label{BoundCondition}
u(x,t) \to 0, \quad x, t \to \pm \infty.
\]

Fundamental (first-order) rogue waves to the Boussinesq equation (\ref{BoussiEq}) were derived in \cite{Tajiri1,Rao2017} by taking a long-wave limit of the two-soliton solution, following a similar procedure introduced in \cite{Ablowitz_Satsuma1978} where several singular rational solutions for the Boussinesq equation (\ref{Eqeta}) were presented. Higher-order rogue waves to the Boussinesq equation (\ref{BoussiEq}) were recently considered by Clarkson and Dowie \cite{ClarksonDowie2017}. By converting this equation into a bilinear one, assuming certain polynomial forms for the bilinear solution, equating powers of $x,t$ in the bilinear equation and solving the resulting algebraic equations for the polynomial coefficients, the authors obtained second- to fifth-order rogue waves with two free real parameters at each order.

Despite this progress, many important questions on rogue waves of the Boussinesq equation are still open. One of the most significant questions is that analytical expressions for rogue waves of arbitrary orders are still unknown. A related question is how many free real parameters exist in general rogue waves of arbitrary orders. Recall that all higher-order rogue waves reported in \cite{ClarksonDowie2017} contained the same number of free real parameters (which is two). But from experience of rogue waves in other integrable equations (such as the NLS equation \cite{DGKM2010,GLML2012,OhtaJY2012}), one anticipates that the number of free parameters should increase with the order of the rogue wave. Thus, general rogue waves of high order in the Boussinesq equation should contain more parameters than two, and those solutions with more free parameters are still unclear. A third open question is what is the maximum amplitude that can be attained in rogue waves of a given order, and what is the spatial-temporal profile of that rogue wave of maximum amplitude. Since the Boussinesq equation is a physically and mathematically interesting equation, these open questions on its rogue waves clearly merit thorough investigation.

In this article, we derive general rogue waves of arbitrary orders in the Boussinesq equation (\ref{BoussiEq}) under boundary conditions (\ref{BoundCondition}). The technique we will use is the bilinear Kadomtsev-Petviashvili (KP) reduction method \cite{Hirota_book}. Although this technique has been applied to derive rogue waves in several other integrable equations before \cite{OhtaJY2012,OhtaJKY2012,OhtaJKY2013,OhtaJKY2014,XiaoeYong2018,Chen_Juntao,SunLian2018,YangYang2019}, when it is applied to the Boussinesq equation, the previous choices of differential operators within this technique would cause considerable technical complications. Here, we propose a new judicious choice of those differential operators, which will drastically simplify the dimension reduction calculation as well as the analytical expressions of rogue wave solutions. Our rogue solutions are given as Gram determinants with $2N-2$ free irreducible real parameters, where $N$ is the order of the rogue wave. Thus, rogue waves of higher orders do contain more free parameters, and the rogue waves reported in \cite{ClarksonDowie2017} are special cases. Tuning these free parameters, we obtain rogue waves of various interesting patterns, many of which have not been seen before. We also find that in the Boussinesq equation, the rogue wave of maximum amplitude at each order is generally \emph{asymmetric} in space. This is a surprising feature which contrasts rogue waves in other integrable equations.

\section{General rogue-wave solutions}
Our general rogue waves in the Boussinesq equation (\ref{BoussiEq}) are given by the following two theorems.

\textbf{Theorem 1.} \, \emph{The Boussinesq equation (\ref{BoussiEq}) under boundary conditions (\ref{BoundCondition}) admits rational nonsingular rogue-wave solutions}
\[  \label{Th1-solutions}
u(x,t)=2 \partial_{x}^2 \ln \sigma ,
\]
\emph{where}
\[ \label{SigmanAlg}
\sigma(x,t)=\det_{
\begin{subarray}{l}
1\leq i, j \leq N
\end{subarray}
}
\left( m_{2i-1,2j-1} \right),
\]
\emph{the matrix elements in $\sigma$ are defined by}
\begin{eqnarray} \label{mijtheorem1}
&& m_{i,j}=\sum_{k=0}^{i} \sum_{l=0}^{j} \frac{a_{k}}{(i-k)!}\frac{(-1)^l a^*_{l}}{(j-l)!}
\left[f(p)\partial_{p}\right]^{i-k}  \left[f(q)\partial_{q}\right]^{j-l} \left. \left(\frac{1}{p+q} e^{\frac{1}{2}(p+q)x-\frac{1}{4}(p^2-q^2)\textrm{i}t}\right)\right|_{p=-1, \ q=-1},
\end{eqnarray}
\emph{with}
\[\label{005-1}
f(p)=\frac{\sqrt{p^2-4}}{3}, \quad  f(q)=\frac{\sqrt{q^2-4}}{3},
\]
\emph{the asterisk `*' represents complex conjugation, and $a_{k}$ are free complex constants.}

The matrix elements in the above theorem are expressed in terms of derivatives with respect to dummy parameters $p$ and $q$. Purely algebraic expressions for these matrix elements can also be obtained through elementary Schur polynomials. For this purpose, we first introduce elementary Schur polynomials $S_k(\mbox{\boldmath $x$})$ which are defined via the generating function,
\[ \label{def_Schur}
\sum_{k=0}^{\infty}S_k(\mbox{\boldmath $x$})\lambda^k
=\exp\left(\sum_{j=1}^{\infty}x_j\lambda^j\right),
\]
where $\mbox{\boldmath $x$}=(x_1,x_2,\cdots)$. Specifically, we have
\[
S_0(\mbox{\boldmath $x$})=1, \quad S_1(\mbox{\boldmath $x$})=x_1,
\quad S_2(\mbox{\boldmath $x$})=\frac{1}{2}x_1^2+x_2, \quad \cdots, \quad
S_{k}(\mbox{\boldmath $x$}) =\sum_{l_{1}+2l_{2}+\cdots+ml_{m}=k} \left( \ \prod _{j=1}^{m} \frac{x_{j}^{l_{j}}}{l_{j}!}\right).
\]
In terms of these Schur polynomials, matrix elements $m_{i,j}$ in Eq. (\ref{mijtheorem1}) can be replaced by formulae in the following theorem.

\textbf{Theorem 2.} \emph{Purely algebraic expressions of matrix elements $m_{i,j}$ in Eq. (\ref{mijtheorem1}) can be written as}
\[ \label{matrixmnij}
m_{i,j}=\sum_{\nu=0}^{\min(i,j)} \Phi_{i\nu} \Psi_{j\nu} ,
\]
\emph{where}
\begin{eqnarray}
&& \Phi_{i\nu} =\left(\frac{\textrm{i}}{2\sqrt{3}}\right)^{\nu}\sum_{k=0}^{i-\nu}a_{k}S_{i-\nu-k}(\textbf{\emph{x}}^{+}+\nu \textbf{\emph{s}}),  \label{phipsiexp1}\\
&& \Psi_{j\nu} =\left(\frac{\textrm{i}}{2\sqrt{3}}\right)^{\nu}\sum_{l=0}^{j-\nu}
(-1)^l a^*_{l} S_{j-\nu-l}(\textbf{\emph{x}}^{-}+\nu \textbf{\emph{s}}),\label{phipsiexp2}
\end{eqnarray}
\emph{and vectors} $\textbf{\emph{x}}^{\pm}=\left(  x_{1}^{\pm}, x_{2}^{\pm},\cdots \right)$, $\emph{\textbf{r}}=(r_{1}, r_{2}, \cdots)$, $\emph{\textbf{s}}=(s_{1}, s_{2}, \cdots)$ \emph{are defined by}
\begin{eqnarray}
&&x_{k}^{\pm}=\frac{e^{2\textrm{i}\pi/3}+(-1)^k e^{-2\textrm{i}\pi/3}}{2\cdot 3^{k}\cdot k!} \hspace{0.03cm} \left[ x\pm (-2)^{k-1} \textrm{i}t \right]+r_{k},  \label{skrkexpcoeff}\\
&& \sum_{k=1}^{\infty} r_{k}\lambda^{k}=-\ln\left[\frac{1}{2}-\cosh\left(\frac{\lambda}{3}+ \frac{2\textrm{i}\pi}{3}\right)\right],\label{skrkexpcoeff1}\\
&& \sum_{k=1}^{\infty} s_{k}\lambda^{k}=\ln\left[ \frac{2\textrm{i}\sqrt{3}}{\lambda} \tanh\frac{\lambda}{6}\tanh\left(\frac{\lambda}{6}+ \frac{2\textrm{i}\pi}{3}\right) \right]. \label{skrkexpcoeff2}
\end{eqnarray}
\emph{The first few $(r_k, s_k)$ values are}
\[
r_1=\frac{i}{2 \sqrt{3}}, \quad r_2=-\frac{5}{72}, \quad r_3=-\frac{i}{54 \sqrt{3}}, \quad s_1=\frac{2 i}{3 \sqrt{3}}, \quad s_2=-\frac{5}{108}, \quad s_3=-\frac{5 i}{243 \sqrt{3}}.
\]

\textbf{Remark 1.} The two expressions (\ref{mijtheorem1}) and (\ref{matrixmnij}) for the matrix element $m_{i,j}$ in Theorems 1 and 2 differ by a constant factor multiplying an exponential of a linear function in $x$ and $t$, but they give the same solution for $u(x,t)$.

\textbf{Remark 2.} Even though the matrix elements in these theorems contain complex numbers, we will show that the determinant (\ref{SigmanAlg}) and the resulting solution $u(x,t)$ are real (see Sec. \ref{sec:reality}).

\textbf{Remark 3.} The rogue waves in Theorem 1 contain $2N$ free complex parameters $a_0, a_1, \dots, a_{2N-1}$, but these free parameters are reducible. First, we can set $a_{0}=1$ without loss of generality. Second, through a determinant manipulation similar to that in \cite{OhtaJY2012}, we can set
\[
a_{2}=a_{4}=a_{6}=\cdots=a_{even}=0
\]
without loss of generality. These $a_0$ and $a_{\mbox{even}}$ values will be taken throughout the rest of this article. Thirdly, with a shifting of $x$ and $t$, we can also normalize $a_{1}=0$ (as was done in \cite{OhtaJY2012}). Thus, the $N$-th order rogue waves in the Boussinesq equation contain $N-1$ free irreducible complex parameters, or $2N-2$ free irreducible real parameters. This number is the same as that for rogue waves in the NLS equation \cite{OhtaJY2012}. Recalling that the second- and higher-order Boussinesq rogue waves reported in \cite{ClarksonDowie2017} contain only two free real parameters, this means that those solutions are special cases of ours in the above theorems. As an example, our third-order rogue waves for the Boussinesq equation contain 4 free irreducible real parameters, which doubles the number of free real parameters in third-order rogue waves of Ref. \cite{ClarksonDowie2017}.

\textbf{Remark 4.} Using the algebraic expressions of matrix elements in Theorem 2, we can show that when $a_1, a_3, \dots, a_{odd}$ are all purely imaginary (and $a_0, a_2, \dots a_{even}$ are taken as in Remark 3), then the rogue solution (\ref{Th1-solutions}) is symmetric in time, i.e.,
\[
u(x, t)=u(x, -t).
\]
The reason is that, if $t\to -t$, then $x_{k}^{+}\to x_{k}^{-}$. In addition, if all $a_{odd}$ are purely imaginary and all $a_{even}$ purely real, then $a_k^*=(-1)^ka_k$. In this case, we see from Eqs. (\ref{matrixmnij})-(\ref{phipsiexp2}) that $m_{ij}(x,-t)=m_{ji}(x,t)$, and thus $u(x,t)=u(x,-t)$.

\section{Analysis of general rogue waves}
\subsection{First- and second-order rogue waves}
Setting $N=1$ and $a_{1}=0$, we get the first-order rogue wave
\begin{eqnarray} \label{u1xt}
u_{1}(x,t)=2 \partial_{x}^2  \ln  \left(\hat{x}^2+t^2+1\right)= \frac{4(1-\hat{x}^2+t^2)}{1+\hat{x}^2+t^2},
\end{eqnarray}
where $\hat{x}= x+1$. This solution matches the one derived earlier in \cite{Tajiri1,Rao2017,ClarksonDowie2017}, and it does not contain any irreducible free parameters.

For second-order rogue waves, we set $N=2$,
\[
a_{1}=0, \quad a_3=\frac{s_{3}+ \textrm{i} r_{3}}{72\sqrt{3}},
\]
where $s_3$ and $r_3$ are free real constants. Then, the corresponding rogue waves in Eq. (\ref{Th1-solutions}) become
\[ \label{u2xt}
u_{2}(x,t)=2 \partial_{x}^2 \ln \sigma_2,
\]
where
\begin{eqnarray}
&&\sigma_2(x,t)=x^6+t^6+3 x^4t^2 +3 x^2t^4 +14 x^5+14 x t^4 +28 x^3t^2 +90 x^4+128 x^2t^2 +22 t^4 +   \nonumber \\
&& \hspace{1.5cm} 324 x^3+316 x t^2 +  648 x^2+360 t^2+648 x+ 324+   \nonumber \\
&& \hspace{1.5cm} 2 r_{3} \left(x^3-3 x t^2+7 x^2-7t^2+16 x+8\right)+2 s_{3} t \left(t^2-3 x^2-14 x-18\right)+  r_{3}^2+s_{3}^2.   \label{sigma2}
\end{eqnarray}
This solution contains two irreducible free real parameters $s_3$ and $r_3$. Recall that the second-order rogue waves derived in \cite{ClarksonDowie2017} also contain two free real parameters. It turns out that the second-order rogue waves in \cite{ClarksonDowie2017} and our solution (\ref{u2xt}) above are equivalent. Indeed, our function $\sigma_2$ in (\ref{sigma2}) can be rewritten as
\[  \label{sigma2form}
\sigma_2\left(x, \hspace{0.02cm} t\right)=\sigma_2^{(s)}(\hat{x},t)+2\alpha t\left(3\hat{x}^2-t^2+\frac{5}{3}\right)+2\beta \hat{x}\left(\hat{x}^2-3t^2-\frac{1}{3}\right)+\alpha^2+\beta^2,
\]
where $\hat{x}=x+\frac{7}{3}$,  $\alpha=-s_3$, $\beta=r_3-\frac{106}{27}$, and
\[ \label{sigma2s}
\sigma_2^{(s)}(x,t)=x^6+\left(3t^2+\frac{25}{3}\right)x^4+\left(3t^4+30t^2-\frac{125}{9}\right)x^2+t^6+\frac{17}{3}t^4+\frac{475}{9}t^2+\frac{625}{9}
\]
is a second-order $(x,t)$-symmetric rogue wave in the Boussinesq equation \cite{ClarksonDowie2017}. The above form of $\sigma_2(x,t)$ matches the generalized second-order rogue waves derived in \cite{ClarksonDowie2017} (except for a space shift).

\subsection{The third-order rogue waves}
To get general third-order rogue waves, we set $N=3$ and $a_1=0$. The corresponding rogue waves (\ref{Th1-solutions}) contain two irreducible free complex parameters $a_{3}$ and $a_{5}$, or four irreducible free real parameters. This number of irreducible free parameters doubles that in the third-order rogue waves of Ref. \cite{ClarksonDowie2017}. Thus, our third-order rogue waves contain many new solutions. For example, two triangular patterns with different orientations as well as two mixed patterns, together with their corresponding $a_3$ and $a_5$ values, are displayed in Fig. 1. When compared to the third-order rogue patterns reported in \cite{ClarksonDowie2017}, we see that the present wave patterns are quite different. It is noted that some of these patterns, such as the triangular ones in the upper row of Fig. 1, look quite similar to those in the local NLS equation \cite{OhtaJY2012}. But the mixed patterns in the lower row of Fig. 1 have not been seen in the NLS equation.
\begin{figure}[htb]
   \begin{center}
   \vspace{-3.5cm}
   \includegraphics[scale=0.350, bb=0 0 385 567]{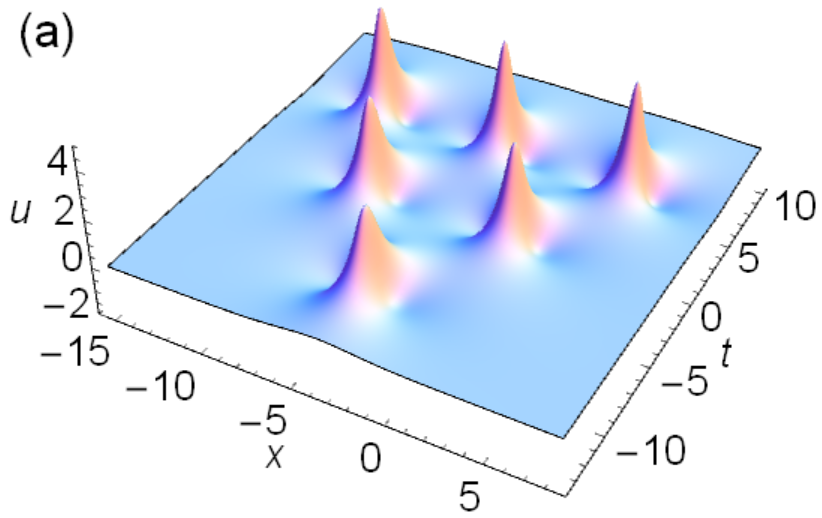}   \hspace{1.5cm}
   \includegraphics[scale=0.350, bb=0 0 385 567]{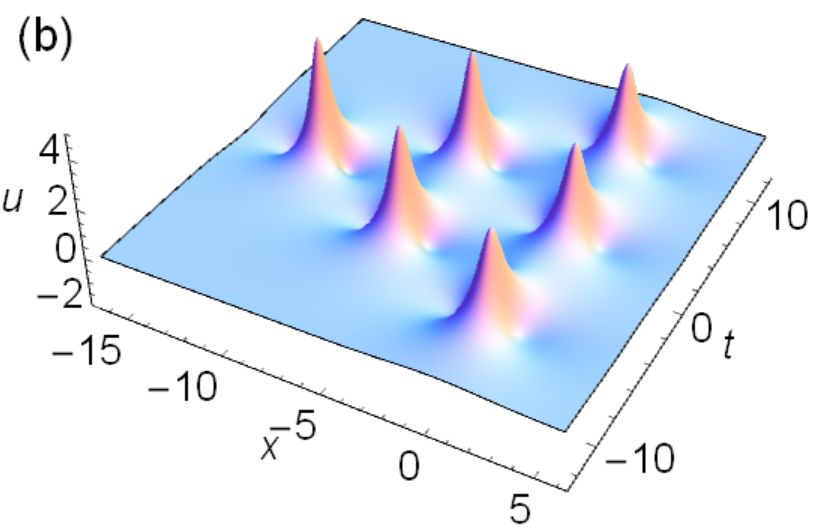}  \\
   \vspace{-3.0cm}
   \includegraphics[scale=0.350, bb=0 0 385 567]{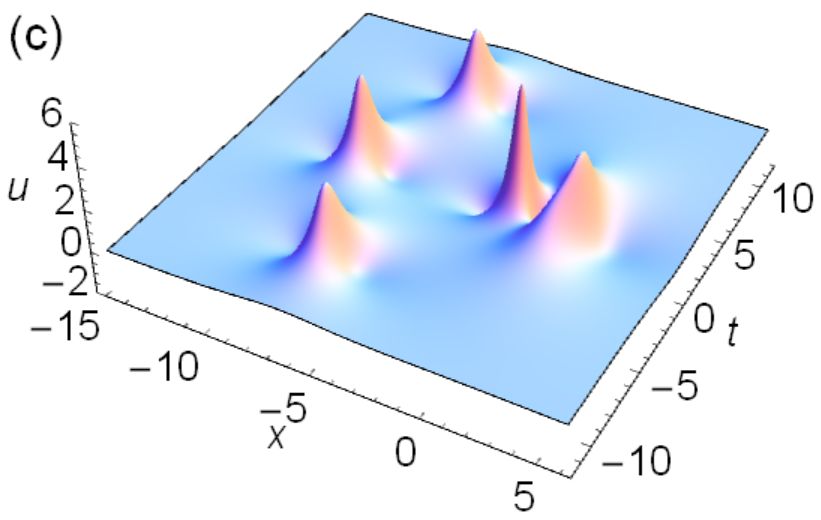}\hspace{1.5cm}
   \includegraphics[scale=0.350, bb=0 0 385 567]{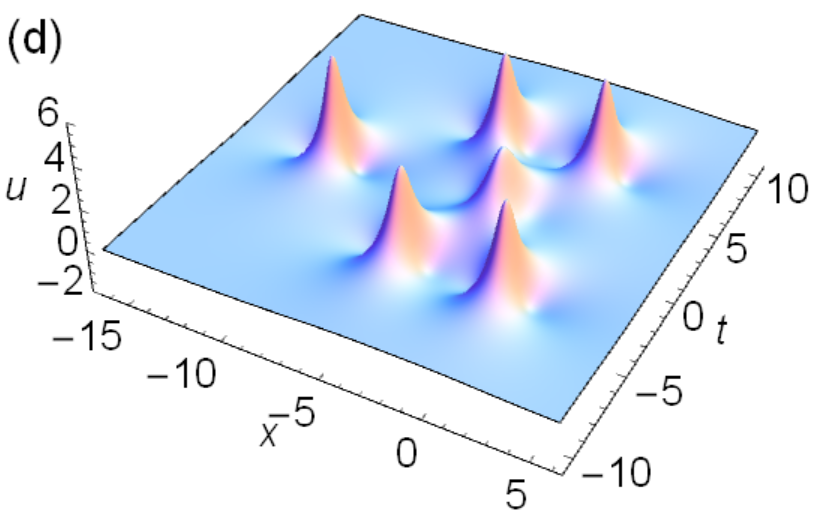}

   \caption{Third-order rogue waves in the Boussinesq equation. Free parameters in these solutions are chosen as
   (a) $a_{3}=1$, $a_{5}=0$; (b) $a_{3}=\textrm{i}$, $a_{5}=0$; (c) $a_{3}=\textrm{i}/2$, $a_{5}=\textrm{i}/2$; and
   (d) $a_{3}=-\textrm{i}/2$, $a_{5}=\textrm{i}/2$.}
   \end{center}
\end{figure}

\subsection{Rogue waves of maximum amplitude}
Next, we explore the maximum amplitude that can be reached by rogue waves of each order. For the NLS equation, this question has been addressed \cite{AAS2009,ACA2010,OhtaJY2012,HeNLSheight}. But for the Boussinesq equation, this question is still open.

For the first-order rogue wave (\ref{u1xt}), since it has no irreducible free parameters, its maximum amplitude can be easily seen as 4, which is attained at $\hat{x}=t=0$. Shifting the location of this maximum amplitude to the origin $(x, t)=(0,0)$, which is equivalent to choosing $a_1=-\textrm{i}/\left(2 \sqrt{3}\right)$ in the rogue waves of Theorems 1 and 2, this first-order rogue wave of maximum amplitude is plotted in Fig. 2(a). Notice that this solution is symmetric in both $x$ and $t$.

For second-order rogue waves, we find that their maximum amplitude is $5.5$. If we require this maximum amplitude to be located at the origin $x=t=0$ (which means that we cannot normalize $a_1$ to be zero), then there are two such rogue waves $u^{(1)}(x,t)$ and $u^{(2)}(x,t)$ with this maximum amplitude, and their corresponding $(a_1, a_3)$ values are
\[ \label{a1a3}
a_1=\textrm{i}\frac{9-7\sqrt{3}}{18}, \quad a_3=\textrm{i}\frac{72-47\sqrt{3}}{324},
\]
and
\[ \label{a1a3b}
a_1=-\textrm{i}\frac{9+7\sqrt{3}}{18}, \quad a_3=-\textrm{i}\frac{72+47\sqrt{3}}{324}.
\]
The profile of the first rogue wave $u^{(1)}(x,t)$ is plotted in Fig. 2(b), and the second wave is related to the first one by $u^{(2)}(x,t)=u^{(1)}(-x,t)$. Notice that these second-order rogue waves of maximum amplitude are symmetric in $t$, but asymmetric in $x$. This contrasts the first-order rogue wave, which is symmetric in both $x$ and $t$ [see Fig. 2(a)].

For third-order rogue waves, their maximum amplitude attained at the origin becomes approximately $6.784$. Two rogue waves feature this maximum amplitude, and they are related to each other by switching $x$ to $-x$. The first wave has parameter values
\[ \label{a1a3a5}
a_{1}\approx -0.1163\textrm{i}, \quad a_{3}\approx -0.0238 \textrm{i}, \quad a_{5}\approx -0.0002 \textrm{i},
\]
and its profile is plotted in Fig. 2(c). The second wave has parameter values
\[
a_{1}\approx -2.001\textrm{i}, \quad a_{3}\approx -2.584 \textrm{i}, \quad a_{5}\approx  -3.937 \textrm{i}.
\]

For fourth-order rogue waves, their maximum amplitude attained at the origin is approximately $7.944$. Again, two rogue waves feature this maximum amplitude. The first one, with parameter values
\[ \label{a1a3a5a7}
a_{1}\approx -0.0799\textrm{i}, \quad a_{3}\approx -0.0213 \textrm{i}, \quad a_{5}\approx 0.0001 \textrm{i}, \quad a_{7}\approx -0.0001 \textrm{i},
\]
is plotted in Fig. 2(d), and the second one has parameter values
\[
a_{1}\approx -1.080\textrm{i}, \quad a_{3}\approx -0.348 \textrm{i}, \quad a_{5}\approx  -0.133 \textrm{i}, \quad a_{7}\approx -0.051\textrm{i}.
\]

For fifth- and sixth-order rogue waves, their maximum amplitudes are approximately 9.017 and 10.025 respectively. Profiles and parameter values for these rogue waves of maximum amplitude are omitted.

\begin{figure}[htb]
   \begin{center}
   \vspace{-3.5cm}
   \includegraphics[scale=0.350, bb=0 0 385 567]{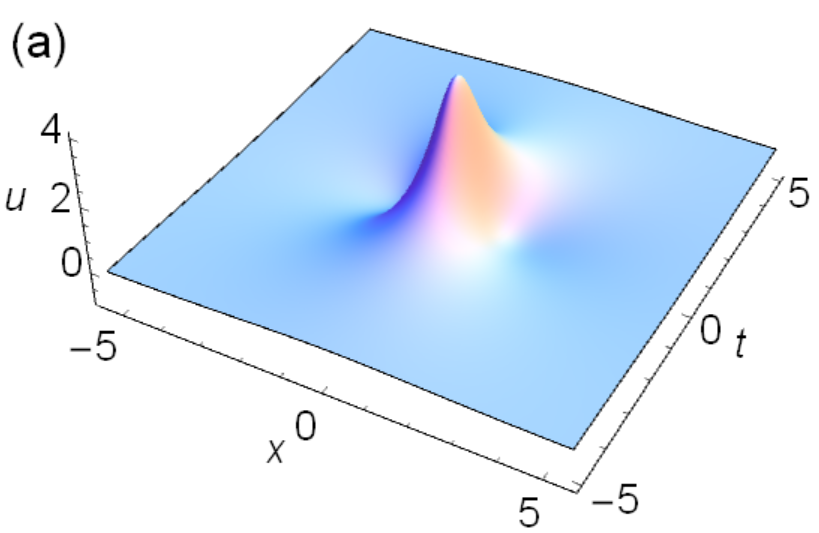}   \hspace{1.5cm}
   \includegraphics[scale=0.350, bb=0 0 385 567]{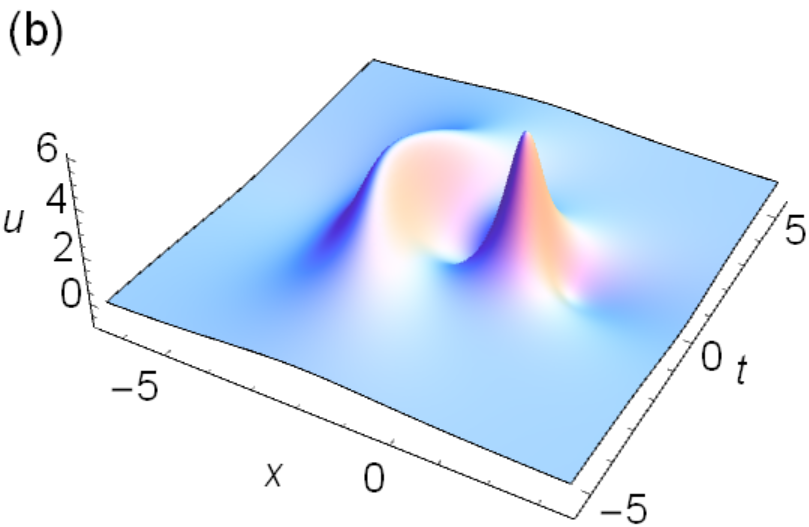}  \\
   \vspace{-3.0cm}
   \includegraphics[scale=0.350, bb=0 0 385 567]{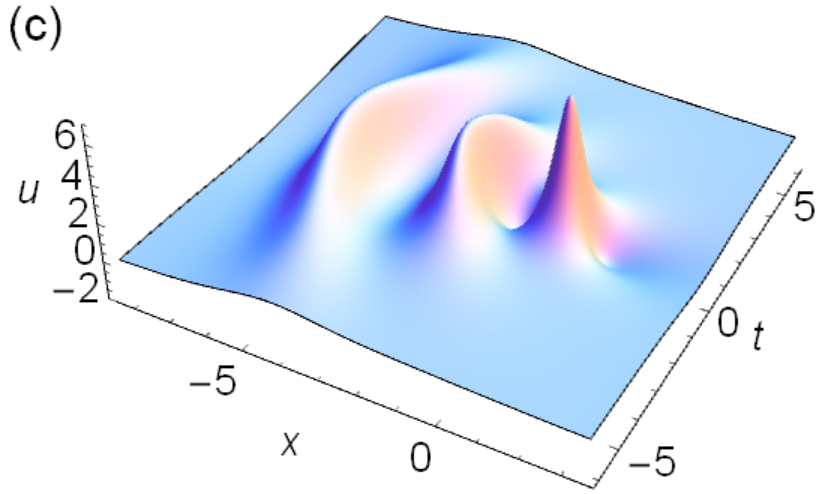}   \hspace{1.5cm}
    \includegraphics[scale=0.350, bb=0 0 385 567]{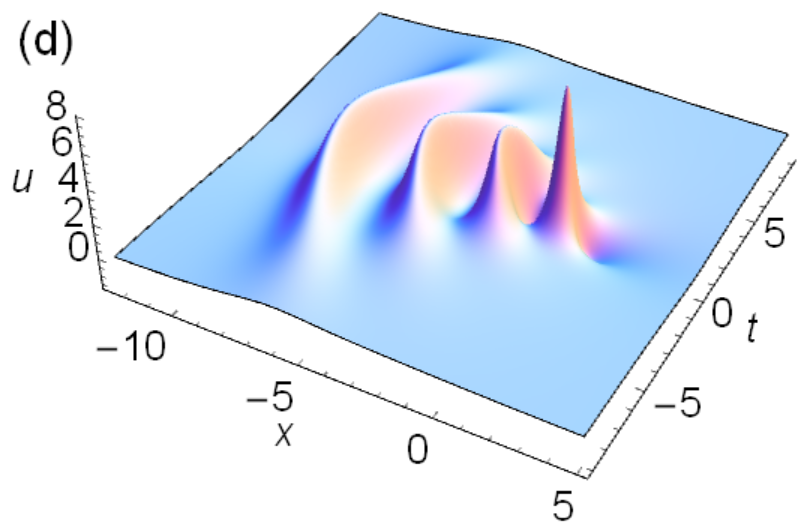}
   \caption{Profiles of the first- to fourth-order rogue waves of maximum amplitude in the Boussinesq equation. The parameter values for these rogue waves are given in the text.}
   \end{center}
\end{figure}

All rogue waves of maximum amplitude from the second order up are asymmetric in $x$ and symmetric in $t$. At each order, two such waves exist which are mirror images of each other around $x=0$.

The above results are briefly summarized in the following table.

\begin{table}[h]
\caption{Maximum amplitudes of rogue waves at each order $N$ and their $x$-symmetry}
\begin{center}
  \begin{tabular}{ | c | c | c|  c | c| }  \hline
         $N$          & Maximum amplitude         & $x$-symmetry  & Amplitude of symmetric wave\\ \hline
          1 &               $4$                   & symmetric     &  4 \\ \hline
          2 &               $5.5$                 & asymmetric    & 4.846 \\ \hline
          3 &               $6.784$             & asymmetric  &  6.545  \\ \hline
          4 &               $7.944$          & asymmetric    & 6.956 \\ \hline
          5 &               $9.017$          & asymmetric    & 8.648 \\ \hline
          6 &               $10.025$          & asymmetric   &  8.757 \\ \hline
  \end{tabular}
\end{center}
\end{table}

It should be noted that at each order, there does exist a rogue wave which is symmetric in both $x$ and $t$. Such symmetric solutions up to order six have been derived and plotted in Ref.~\cite{ClarksonDowie2017}. They are special solutions of our general rogue waves (\ref{Th1-solutions}) as well. For instance, the first-order symmetric rogue wave is obtained from our formula (\ref{Th1-solutions}) when we choose $a_{1}=-\textrm{i}/(2 \sqrt{3})$; the second-order one is obtained when we choose $a_{1}=-7\textrm{i}/(6\sqrt{3})$ and $a_{3}=-31\textrm{i}/(216 \sqrt{3})$; the third-order one is obtained when we choose $a_{1}=-11\textrm{i}/(6\sqrt{3})$, $a_{3}=-409\textrm{i}/(648 \sqrt{3})$ and $a_{5}=-1619\textrm{i}/(6480 \sqrt{3})$; and for the fourth-order, the parameters are $a_{1}=-5\textrm{i}/(2\sqrt{3})$, $a_{3}=-359\textrm{i}/(216 \sqrt{3})$, $a_{5}=-5041\textrm{i}/(3888 \sqrt{3})$, and $a_{7}=-29937263\textrm{i}/(29393280  \sqrt{3})$; and so on.

However, except for order one, this symmetric rogue wave does not attain the maximum amplitude of that order. For instance, the peak amplitudes of symmetric rogue waves from the second to sixth  orders are approximately 4.846, 6.545, 6.956, 8.648 and 8.757 respectively. These peak values of symmetric rogue waves are also listed in Table 1. Compared to the maximum amplitudes in that table, it is clear that these peak amplitudes of symmetric rogue waves are lower than the maximum amplitudes of $x$-asymmetric rogue waves (except for order one).

The fact that rogue waves of maximum amplitude in the Boussinesq equation are generally asymmetric rather than symmetric in space is very counterintuitive, considering that the Boussinesq equation itself is symmetric in space. This asymmetry also contrasts the NLS equation, where the maximum-amplitude rogue waves are always symmetric in space \cite{AAS2009,ACA2010,OhtaJY2012,HeNLSheight}. These asymmetric wave patterns, as shown in Fig. 2(b-d), are very novel and have not been seen in rogue waves of other integrable equations (to the authors' best knowledge).

\section{Derivation of rogue-wave solutions}
In this section, we derive the general rogue-wave solutions given in Theorems 1 and 2. This derivation uses the bilinear KP-reduction method in the soliton theory. This method is based on Hirota's bilinear form of an integrable equation \cite{Hirota_book}, and the observation that this bilinear equation is often a member of the KP hierarchy \cite{Jimbo_Miwa} (possibly after certain reductions such as the dimension reduction). Thus, solutions of the KP hierarchy, under restrictions due to those reductions, will provide solutions to the original integrable system. This technique often gives elegant determinant-type solutions. In addition, it produces solutions to higher-dimensional integrable equations more easily than to lower-dimensional ones, which is remarkable. This bilinear KP-reduction technique has been applied to derive rogue waves in several other integrable equations before \cite{OhtaJY2012,OhtaJKY2012,OhtaJKY2013,OhtaJKY2014,XiaoeYong2018,Chen_Juntao,SunLian2018,YangYang2019}. Since rogue waves are rational solutions, the way to derive such rational terms in this bilinear KP method is to define matrix elements as certain differential operators (with respect to parameters) acting on an exponential term. When this technique is applied to the Boussinesq equation, however, the previous choices of such differential operators would cause considerable technical difficulties for the dimension reduction (see \cite{XiaoeYong2018,Chen_Juntao} for examples). To overcome those difficulties, we will introduce a new and general way to choose these differential operators. As we will see, this new treatment will streamline the dimension reduction calculation and simplify the analytical expressions of rogue wave solutions.

The outline of our derivation is as follows.

First, we make the standard variable transformation
\[\label{usolution}
u(x,t)=2\partial_x^2 \ln \sigma(x,t),
\]
where $\sigma(x,t)$ is a real variable.  Under this transformation, the Boussinesq equation (\ref{BoussiEq}) is converted into the bilinear form \cite{ClarksonDowie2017}
\begin{eqnarray}\label{Bilineform1}
    \left( D_{x}^4- 3 D_{x}^2-3 D_{t}^{2} \right) \sigma \cdot \sigma = 0,
\end{eqnarray}
where $D$ is Hirota's bilinear differential operator defined by
\begin{eqnarray*}
&&P \left(D_{x}, D_{y}, D_{t},\cdots\right) F(x,y,t,\cdots) \cdot G(x,y,t,\cdots) \\
&& \equiv P\left(\partial_{x}-\partial_{x'}, \partial_{y}-\partial_{y'}, \partial_{t}-\partial_{t'}, \cdots \right) F(x,y,t,\cdots) G(x',y',t',\cdots)|_{x'=x, y'=y, t'=t,\cdots},
\end{eqnarray*}
and $P$ is a polynomial of $D_{x}$, $D_{y}$, $D_{t}, \cdots .$

In order to derive solutions to the bilinear equation (\ref{Bilineform1}), we consider a higher-dimensional bilinear equation
\begin{eqnarray}\label{2DBilineform}
\left( D_{x_{1}}^4- 4 D_{x_{1}}D_{x_{3}}+3 D_{x_{2}}^{2} \right) \sigma \cdot \sigma = 0,
\end{eqnarray}
which is the bilinear form of the KP equation \cite{Hirota_book,Jimbo_Miwa}. We first construct a wide class of algebraic solutions for this higher-dimensional bilinear equation in the form of Gram determinants. Then, we restrict these solutions so that they
satisfy the dimension-reduction condition
\[\label{DimentionPara}
\left(\partial_{x_3}-3\partial_{x_1}\right) \sigma=C\sigma,
\]
where $C$ is some constant (the reason for our choice of the coefficient $3$ in front of $\partial_{x_1}$ will be explained later). Under this condition, the higher-dimensional bilinear equation (\ref{2DBilineform}) reduces to
\begin{eqnarray}\label{2DBilineform2}
\left( D_{x_{1}}^4- 12 D_{x_{1}}^2+3 D_{x_{2}}^{2} \right) \sigma \cdot \sigma = 0.
\end{eqnarray}
Finally, we define
\[ \label{x1x2}
x_1=\frac{1}{2}x, \quad x_2=-\frac{1}{4}\textrm{i}t,
\]
and impose the reality condition
\[\label{NonlocPara}
\sigma^*=\sigma.
\]
Then, the bilinear equation (\ref{2DBilineform2}) becomes the bilinear equation (\ref{Bilineform1}) of the Boussinesq equation, and $\sigma$ (with $x_3=0$) becomes its algebraic (rogue wave) solution.

Next, we follow the above outline to derive general rogue-wave solutions to the Boussinesq equation (\ref{BoussiEq}).

\subsection{Gram solutions for the higher-dimensional bilinear system}
First, we derive algebraic solutions to the higher-dimensional bilinear equation (\ref{2DBilineform}). From Lemma 1 of Ref.~\cite{GmuZQin2014}, we learn that if functions $m_{i,j}^{(n)}$, $\varphi_{i}^{(n)}$ and $\psi_{j}^{(n)}$ of variables ($x_{1}$, $x_{2}$, $x_{3}$) satisfy the following differential and difference relations
\begin{eqnarray}\label{003a}
\left.
  \begin{array}{ll}
  \partial_{x_{1}}m_{i,j}^{(n)}=\varphi_{i}^{(n)}\psi_{j}^{(n)},\\
  \partial_{x_{2}}m_{i,j}^{(n)}=\varphi_{i}^{(n+1)}\psi_{j}^{(n)}+\varphi_{i}^{(n)}\psi_{j}^{(n-1)},\\
  \partial_{x_{3}}m_{i,j}^{(n)}=\varphi_{i}^{(n+2)}\psi_{j}^{(n)}+\varphi_{i}^{(n+1)}\psi_{j}^{(n-1)}+\varphi_{i}^{(n)}\psi_{j}^{(n-2)},\\
  m_{i,j}^{(n+1)}=m_{i,j}^{(n)}+\varphi_{i}^{(n)}\psi_{j}^{(n+1)},\\
  \partial_{x_{k}}\varphi_{i}^{(n)}=\varphi_{i}^{(n+k)},\  \partial_{x_{k}}\psi_{j}^{(n)}=-\psi_{j}^{(n-k)}, \ (k=1,2,3),
  \end{array}
\right\}
\end{eqnarray}
then the determinant
\[\label{Mijdeterminants}
\tau_{n}=\det_{1\leq i,j \leq N} \left(m_{i,j}^{(n)}\right)
\]
satisfies the bilinear equation (\ref{2DBilineform}), i.e.,
\begin{eqnarray}\label{BiliEqLemma}
 \left( D_{x_{1}}^4- 4 D_{x_{1}}D_{x_{3}}+3 D_{x_{2}}^{2} \right) \tau_{n} \cdot \tau_{n} = 0.
\end{eqnarray}

Next, we introduce functions $m^{(n)}$, $\varphi^{(n)}$ and $\psi^{(n)}$ as
\[ \label{mn}
m^{(n)}=\frac{1}{p+q}\left(-\frac{p}{q}\right)^ne^{\xi+\eta}, \quad
\varphi^{(n)}=p^ne^{\xi}, \quad
\psi^{(n)}=(-q)^{-n}e^{\eta},
\]
where
\[
\xi= p x_{1}+p^2x_{2}+p^3x_{3}, \quad \eta=q x_{1}-q^2x_{2}+q^3x_{3}. \label{002-2}
\]
It is easy to check that these functions satisfy the differential and difference relations
\begin{eqnarray*}
&&\partial_{x_1} m^{(n)}=\varphi^{(n)}\psi^{(n)},
\\
&&\partial_{x_2} m^{(n)}
=\varphi^{(n+1)}\psi^{(n)}+\varphi^{(n)}\psi^{(n-1)},
\\
&&\partial_{x_{3}} m^{(n)}=\varphi^{(n+2)}\psi^{(n)}+\varphi^{(n+1)}\psi^{(n-1)}+\varphi^{(n)}\psi^{(n-2)},
\\
&& m^{(n+1)}=m^{(n)}+\varphi^{(n)}\psi^{(n+1)},
\\
&&\partial_{x_k}\varphi^{(n)}=\varphi^{(n+k)}, \quad
\partial_{x_k}\psi^{(n)}=-\psi^{(n-k)},
\quad (k=1,2,3).
\end{eqnarray*}
Therefore, by defining
\[
m_{ij}^{(n)}=\mathcal{A}_i \mathcal{B}_{j} m^{(n)}, \quad
\varphi_i^{(n)}=\mathcal{A}_i\varphi^{(n)}, \quad
\psi_j^{(n)}=\mathcal{B}_{j}\psi^{(n)},    \label{mijn}
\]
where $\mathcal{A}_{i}$ and $\mathcal{B}_{j}$ are differential operators with respect to $p$ and $q$ respectively as
\begin{eqnarray}\label{003b}
\left.
  \begin{array}{ll}
    \mathcal{A}_{i}=\sum_{k=0}^{i}\frac{a_{k}}{(i-k)!}\left[f_1(p)\partial_{p}\right]^{i-k} \vspace{0.25cm}\\
 \mathcal{B}_{j}=\sum_{l=0}^{j}\frac{b_{l}}{(j-l)!}\left[f_2(q)\partial_{q}\right]^{j-l}
  \end{array}
\right\},
\end{eqnarray}
$f_1(p)$ and $f_2(q)$ are arbitrary functions of $p$ and $q$, and $a_{k}$, $b_{k}$ are arbitrary complex constants, we would see that these $ m_{ij}^{(n)}$, $\varphi_i^{(n)}$ and $\psi_j^{(n)}$ obey the differential and difference relations
(\ref{003a}) since the operators $\mathcal{A}_{i}$ and $\mathcal{B}_{j}$ commute with
differentials $\partial_{x_k}$. Then, Lemma 1 of Ref.~\cite{GmuZQin2014} tells us that for an
arbitrary sequence of indices $(i_1,i_2,\cdots,i_N;j_1,j_2,\cdots,j_N)$, the determinant
\[ \label{tildetaun}
\tau_n=\det_{1\le\nu,\mu\le N}\left( m_{i_\nu,j_\mu}^{(n)}\right)
\]
satisfies the higher-dimensional bilinear equation (\ref{BiliEqLemma}).

The above solutions (\ref{tildetaun}) are a very broad class of algebraic solutions to the bilinear equation (\ref{BiliEqLemma}) which contain a huge amount of freedom. For instance, parameters $p$ and $q$ are totally arbitrary, so are the functions $f_1(p)$ and $f_2(q)$, complex constants $a_{k}$ and $b_{k}$, as well as sequences of indices $(i_1,i_2,\cdots,i_N;j_1,j_2,\cdots,j_N)$. Only a small portion of such solutions can satisfy the dimension reduction condition (\ref{DimentionPara}) and the reality condition (\ref{NonlocPara}), which then make them algebraic solutions to the bilinear equation (\ref{Bilineform1}) of the the Boussinesq equation (\ref{BoussiEq}) under variable connections (\ref{x1x2}).

Next, we will restrict solutions (\ref{tildetaun}) so that they meet the dimension reduction condition (\ref{DimentionPara}) and the reality condition (\ref{NonlocPara}).

\subsection{Dimensional reduction through the $\mathcal{W}$-$p$ treatment}
We first condider the dimensional reduction (\ref{DimentionPara}) for the bilinear equation (\ref{BiliEqLemma}), which will introduce restrictions on parameters $p$ and $q$, functions $f_1(p)$ and $f_2(q)$, as well as sequences of indices $(i_1,i_2,\cdots,i_N;j_1,j_2,\cdots,j_N)$. We will also explain the choice of the coefficient $3$ in front of $\partial_{x_1}$ in that reduction (\ref{DimentionPara}).

Dimension reduction is a crucial step in the bilinear KP-reduction procedure. In the past, the functions $f_1(p)$ and $f_2(q)$ in this procedure were always chosen to be linear functions \cite{OhtaJY2012,OhtaJKY2012,OhtaJKY2013,OhtaJKY2014,XiaoeYong2018,Chen_Juntao,SunLian2018,YangYang2019}.
In many cases, such choices were appropriate. However, for a coupled NLS-Boussinesq system and the Yajima-Oikawa system studied in \cite{XiaoeYong2018,Chen_Juntao}, dimension reduction under such choices became cumbersome and complicated. Indeed, the authors were forced to introduce two additional indices in the coefficients of the differential operators and matrix elements, and those coefficients had to be obtained from nontrivial recurrence relations. For the present Boussinesq equation (\ref{BoussiEq}), choices of linear functions for $f_1(p)$ and $f_2(q)$ would encounter a similar difficulty as in \cite{XiaoeYong2018,Chen_Juntao}.

In this article, we will choose functions $f_1(p)$ and $f_2(q)$ differently. We will show that under our judicious choices of $f_1(p)$ and $f_2(q)$, dimension reduction will simplify dramatically. As a result, recurrence relations for coefficients in the solution will be eliminated, and the solution expressions of rogue waves will become more clean and concise.

We start from a general dimension reduction condition
\[ \label{dx3alphax1}
\left(\partial_{x_3}+\alpha \partial_{x_1}\right) \tau_n=C\tau_n,
\]
where $\alpha$ and $C$ are constants to be determined. In order to calculate the left side of the above equation, we notice from the definitions (\ref{mn}) and (\ref{mijn}) of $m^{(n)}$ and $m_{i,j}^{(n)}$ that
\[ \label{x3x1mij}
\left(\partial_{x_3}+\alpha \partial_{x_1}\right) m_{i,j}^{(n)} = \mathcal{A}_{i}\mathcal{B}_{j} \left[Q_1(p)+Q_2(q)\right] m^{(n)},
\]
where
\[ \label{Qpdef}
Q_1(p)=p^3+ \alpha p, \quad Q_2(q)=q^3+ \alpha q.
\]
To proceed, we introduce new variables $\mathcal{W}_1$ and $\mathcal{W}_2$ through
\begin{eqnarray}\label{QpfunctionWp}
Q_1(p)=\mathcal{W}_1(p)+\frac{1}{\mathcal{W}_1(p)}, \quad Q_2(q)=\mathcal{W}_2(q)+\frac{1}{\mathcal{W}_2(q)}.
\end{eqnarray}
In terms of these new variables, our new choices of functions $f_1(p)$ and $f_2(q)$ in the differential operators $\mathcal{A}_{i}$ and $\mathcal{B}_{j}$ are
\begin{eqnarray} \label{f1pf2q}
f_1(p)=\frac{\mathcal{W}_1(p)}{\mathcal{W}'_1(p)}, \quad f_2(q)=\frac{\mathcal{W}_2(q)}{\mathcal{W}'_2(q)}.
\end{eqnarray}
More explicit expressions for $f_1(p)$ and $f_2(q)$ will be provided later after the value of $\alpha$ is ascertained [see Eq. (\ref{005-1b})].
The motivation behind this choice of $f_1(p)$ is that, under this choice,
\[ \label{f1pmot}
f_1(p)\partial_{p}=\frac{\mathcal{W}_1(p)}{\mathcal{W}'_1(p)}\partial_{p}=\mathcal{W}_1(p)\partial_{\mathcal{W}_1(p)}.
\]
Thus,
\[
\mathcal{A}_{i}Q_1(p)m^{(n)}=\left[\sum_{k=0}^{i}\frac{a_{k}}{(i-k)!}\left(\mathcal{W}_1\partial_{\mathcal{W}_1}\right)^{i-k}\right]\left(\mathcal{W}_1+\frac{1}{\mathcal{W}_1}\right)m^{(n)}.
\]
One can recognize that the operators on the right side of this equation are the same as those in \cite{OhtaJY2012}, except for a notation of $\mathcal{W}_1$ instead of $p$. Using results of \cite{OhtaJY2012}, we immediately get
\begin{eqnarray}
\mathcal{A}_{i}Q_1(p)m^{(n)}= \sum_{k=0}^i \frac{1}{k!} \left[ \mathcal{W}_1(p)+(-1)^k \frac{1}{\mathcal{W}_1(p)} \right] \mathcal{A}_{i-k}m^{(n)}.
\end{eqnarray}
For exactly the same reasons, we also have
\begin{eqnarray}
\mathcal{B}_{j} Q_2(q) m^{(n)}= \sum_{l=0}^j \frac{1}{l!} \left[ \mathcal{W}_2(q)+(-1)^l \frac{1}{\mathcal{W}_2(q)} \right] \mathcal{B}_{j-l}m^{(n)}.
\end{eqnarray}
Substituting these two equations into (\ref{x3x1mij}), we then get
\begin{eqnarray} \label{dx3x1mij}
\left(\partial_{x_3}+\alpha \partial_{x_1}\right) m_{i,j}^{(n)}=
\sum_{k=0}^i \frac{1}{k!} \left[ \mathcal{W}_1(p)+(-1)^k \frac{1}{\mathcal{W}_1(p)} \right] m_{i-k,j}^{(n)}+\sum_{l=0}^j \frac{1}{l!} \left[ \mathcal{W}_2(q)+(-1)^l \frac{1}{\mathcal{W}_2(q)} \right] m_{i,j-l}^{(n)}.
\end{eqnarray}

Now, it is time to select values of $p$, $q$ and $\alpha$ so that the above equation can be further simplified (the selected values of $p$ and $q$ will be denoted as $p_0$ and $q_0$). Since treatments for $p_0$ and $q_0$ values are the same, we will consider $p_0$ only. Motivated by \cite{OhtaJY2012}, we require $W_1(p_0)=1$, so that the odd-$k$ terms in the above summation drop out. This $W_1(p_0)=1$ condition leads to $Q_1(p_0)=2$ in view of Eq. (\ref{QpfunctionWp}). Differentiating the first equation in (\ref{QpfunctionWp}) with respect to $p$, we get
\[
\mathcal{W}'_1(p)=\frac{Q'_1(p)}{1-\mathcal{W}_1^{-2}(p)}.
\]
At the selected $p_0$ value where $W_1(p_0)=1$, $\mathcal{W}'_1(p_0)$ needs to be well defined in view of the definition of $f_1(p)$ in
Eq. (\ref{f1pf2q}). Then, the above equation requires $Q'_1(p_0)=0$. Thus, the two conditions for the $p_0$ and $\alpha$ values are
\[
Q_1(p_0)=2, \quad Q'_1(p_0)=0.
\]
Inserting the $Q_1(p)$ function from (\ref{Qpdef}) into these two constraints, we get
\[
p_0^3+\alpha p_0=2, \quad 3p_0^2+\alpha=0,
\]
whose solutions are
\[ \label{p0alpha}
p_0=-1, \quad \alpha=-3.
\]
This explains the coefficient $3$ we have chosen in the dimension reduction condition (\ref{DimentionPara}). The same consideration for $q_0$ leads to $q_0=-1$.

Under the above choices of $p, q$ and $\alpha$ values, Eq. (\ref{dx3x1mij}) further simplifies to
\[  \label{contigurelati}
\left(\partial_{x_3}-3\partial_{x_1}\right) \left. m_{i,j}^{(n)}\right|_{p=-1, \ q=-1}=2\sum^i_{\begin{subarray}{c} k=0,\\ k: even \end{subarray}}\frac{1}{k!} \left. m_{i-k,j}^{(n)}\right|_{p=-1, \ q=-1}+2\sum_{\begin{subarray}{c} l=0,\\ l: even \end{subarray}}^{j}\frac{1}{l!} \left. m_{i,j-l}^{(n)}\right|_{p=-1, \ q=-1}.
\]
This is an important relation which shows that, at the selected $(p, q)$ values, $\left(\partial_{x_3}-3\partial_{x_1}\right) m_{i,j}^{(n)}$
is a linear combination of $m_{i,j}^{(n)}$ and other $m_{\hat{i},\hat{j}}^{(n)}$ terms of lower row and column indices with jumps of 2. Due to this relation, if we choose indices $(i_1,i_2,\cdots,i_N;j_1,j_2,\cdots,j_N)$ in the determinant (\ref{tildetaun}) as
\begin{eqnarray}\label{Odd-Det-Solution}
\tau_{n}=\det_{1\leq i, j\leq N}\left(m_{2i-1, 2j-1}^{(n)} \right),
\end{eqnarray}
then the same calculation as in \cite{OhtaJY2012} would show that this $\tau_{n}$ function satisfies the dimension reduction condition
\[\label{ReductionCondi}
\left(\partial_{x_{3}}-3 \partial_{x_{1}}\right) \tau_{n} = 4N \tau_{n}.
\]
We note by passing that another index choice of
\begin{eqnarray}\label{Even-Det-Solution}
\tau_{n}=\det_{1\leq i, j\leq N}\left(m_{2i-2, 2j-2}^{(n)} \right)
\end{eqnarray}
would also satisfy the dimension reduction condition (\ref{ReductionCondi}). But as we have shown in a different but similar context \cite{YangYang2019}, this other index choice would lead to solutions which are equivalent to those from (\ref{Odd-Det-Solution}).

When the dimension reduction condition (\ref{ReductionCondi}) is substituted into the higher-dimensional bilinear equation (\ref{BiliEqLemma}) and setting $n=0$ and $x_3=0$, we get
\begin{eqnarray} \label{Dx1tau0}
\left( D_{x_{1}}^4- 12 D_{x_{1}}^2+3 D_{x_{2}}^{2} \right) \tau_{0} \cdot \tau_{0} = 0.
\end{eqnarray}
Thus, the third dimension $x_3$ has been eliminated, and dimension reduction has been completed.

During the above dimension reduction, the $\alpha$ value in (\ref{dx3alphax1}) has been ascertained [see (\ref{p0alpha})]. Thus, we can now derive more explicit expressions for functions $f_1(p)$ and $f_2(q)$ as defined in Eq. (\ref{f1pf2q}). From Eq. (\ref{QpfunctionWp}), we have
\[
\left[\mathcal{W}_1(p)-\frac{1}{\mathcal{W}_1(p)}\right]^2=Q_1^2(p)-4.
\]
Differentiating the first equation in (\ref{QpfunctionWp}) with respect to $p$, we get
\[
\frac{\mathcal{W}'_1(p)}{\mathcal{W}_1(p)}\left[\mathcal{W}_1(p)-\frac{1}{\mathcal{W}_1(p)}\right]=Q'_1(p).
\]
Utilizing these two equations and the definition of $f_1(p)$ in (\ref{f1pf2q}), we get
\[
f_1(p)=\frac{\sqrt{Q_1^2(p)-4}}{Q'_1(p)}.
\]
In view of the definition of $Q_1(p)$ in (\ref{Qpdef}) and the $\alpha$ value in (\ref{dx3alphax1}), the above $f_1(p)$ formula can be further simplified. Following the same calculation, the $f_2(q)$ function can also be derived. The final results are
\[\label{005-1b}
f_1(p)=\frac{\sqrt{p^2-4}}{3},\quad f_2(q)=\frac{\sqrt{q^2-4}}{3}.
\]

\subsection{The reality condition} \label{sec:reality}

In the bilinear equation (\ref{Dx1tau0}), when the $(x_1, x_2)$ variables are linked to $(x, t)$ through (\ref{x1x2}), i.e., $x_1=x/2$ and $x_2=-\textrm{i}t/4$, then this bilinear equation becomes (\ref{Bilineform1}) of the Boussinesq equation. The only remaining condition is the reality condition (\ref{NonlocPara}), i.e.,
\[\label{ComplConjuTaunf}
\tau^*_{0}=\tau_{0}.
\]
Notice that when $p=q=-1$, $f_1(p)$ and $f_2(q)$ are purely imaginary in view of Eq. (\ref{005-1b}). Hence,
\[
\left(\left[f_1(p)\partial_{p}\right]^{k}\right)^*=(-1)^{k}\left[f_1(p)\partial_{p}\right]^{k}, \quad
\left(\left[f_2(q)\partial_{q}\right]^{l}\right)^*=(-1)^{l}\left[f_2(q)\partial_{q}\right]^{l}.
\]
Thus, if we constrain parameters $a_{k}$ and $b_{k}$ by
\[ \label{CompConguCondi}
b_{k} = (-1)^k a^*_{k},
\]
then since $x_1=x/2$ is real and $x_2=-\textrm{i}t/4$ is imaginary, we can easily show that
\[ \label{mijsym}
\left[m_{i,j}^{(0)}\right]^*= (-1)^{i+j} m_{j,i}^{(0)},
\]
and therefore the reality condition (\ref{ComplConjuTaunf}) holds.

Regarding the regularity of solutions given in Theorem 1, notice from Eqs. (\ref{SigmanAlg}) and (\ref{mijsym}) that $\sigma(x,t)$ is the determinant of a Hermitian matrix $M=\mbox{mat}_{1\le i,j\le N}(m_{2i-1, \hspace{0.04cm} 2j-1})$. Then, using techniques similar to that in \cite{OhtaJY2012}, we can show that the matrix $M$ is positive definite, so that $\sigma(x,t)> 0$, which proves that the solution $u(x,t)$ is nonsingular.

Summarizing the above results, Theorem 1 is then proved, except for the boundary conditions (\ref{BoundCondition}), which will be discussed in the end of the next subsection.

\subsection{Algebraic representation of rogue waves}
Finally, we derive purely algebraic expressions of rogue waves and prove Theorem 2. These algebraic expressions are useful for multiple purposes. For instance, they can produce explicit formulae of rogue waves much more quickly than the expressions in Theorem 1 where repeated differentiations have to be performed. For another instance, these algebraic expressions allow us to derive highest-power polynomial terms of rogue waves \cite{OhtaJY2012,YangYang2018,YangYang2019}, which can be used to analytically prove the boundary conditions (\ref{BoundCondition}) of these solutions.

The basic idea of this derivation is the same as that in \cite{OhtaJY2012}, except that we will work with variables $\mathcal{W}_1(p)$ and $\mathcal{W}_2(q)$ instead of $p$ and $q$. Introducing the generator $\mathcal{G}$ of the differential operators $\left(f_1(p)\partial_{p}\right)^{k} \left(f_2(q)\partial_{q}\right)^{l}$ as
\[ \label{GeneratorG}
\mathcal{G}= \sum_{k=0}^\infty \sum_{l=0}^{\infty} \frac{\kappa^k}{k!} \frac{\lambda^l}{l!} \left[f_1(p)\partial_{p}\right]^{k}  \left[f_2(q)\partial_{q}\right]^{l},
\]
and utilizing Eq. (\ref{f1pmot}), we get
\[ \label{GeneratorG2}
\mathcal{G}= \sum_{k=0}^\infty \sum_{l=0}^{\infty} \frac{\kappa^k}{k!} \frac{\lambda^l}{l!} \left[\partial_{\ln\mathcal{W}_1}\right]^{k}  \left[\partial_{\ln\mathcal{W}_2}\right]^{l} = \exp\left(\kappa \partial_{\ln\mathcal{W}_1} + \lambda \partial_{\ln\mathcal{W}_2}\right).
\]
Thus, for any function $F(\mathcal{W}_1,\mathcal{W}_2)$, we have \cite{OhtaJY2012}
\[ \label{GF}
\mathcal{G} F(\mathcal{W}_1,\mathcal{W}_2)=F(e^{\kappa}\mathcal{W}_1,e^{\lambda}\mathcal{W}_2).
\]

Next, we will apply this generator on the function $m^{(0)}$. After dimension reduction ($x_3=0$) and variable relations (\ref{x1x2}), this $m^{(0)}$ reduces from (\ref{mn}) to
\[
m^{(0)}=\frac{1}{p+q} e^{\frac{1}{2}(p+q)x-\frac{1}{4}(p^2-q^2)\textrm{i}t}.
\]
To utilize Eq. (\ref{GF}), we need to express $p$ and $q$ in this $m^{(0)}$ as functions of $\mathcal{W}_1$ and $\mathcal{W}_2$. Equations (\ref{Qpdef})-(\ref{QpfunctionWp}) tell us that
\begin{eqnarray}
p^3-3p=\mathcal{W}_1+\frac{1}{\mathcal{W}_1}, \quad q^3-3q=\mathcal{W}_2+\frac{1}{\mathcal{W}_2}.  \label{CubicSolu2}
\end{eqnarray}
These equations for $p$ and $q$ can be solved and there are three roots. Due to our earlier conditions that $\mathcal{W}_1=\mathcal{W}_2=1$ when $p=q=-1$, the suitable roots for $p$ and $q$ are
\begin{eqnarray}
p(\mathcal{W}_1)=c_{1}\mathcal{W}_1^{1/3}+ c_{2} \mathcal{W}_1^{-1/3}, \quad
q(\mathcal{W}_2)=c_{1}\mathcal{W}_2^{1/3}+ c_{2} \mathcal{W}_2^{-1/3}, \label{CubicEqRoot2}
\end{eqnarray}
where $c_{1}=\exp\left(2\textrm{i}\pi/3\right)$ and $c_{2}=c_1^*$.

Now, we apply Eq. (\ref{GF}) on $m^{(0)}$ and get
\[
\mathcal{G} m^{(0)} = \frac{1}{p(e^{\kappa}\mathcal{W}_1)+q(e^{\lambda}\mathcal{W}_2)} \exp\left({\frac{1}{2}\left[p(e^{\kappa}\mathcal{W}_1)+q(e^{\lambda}\mathcal{W}_2)\right]x-\frac{1}{4}\left[p^2(e^{\kappa}\mathcal{W}_1)-q^2(e^{\lambda}\mathcal{W}_2)\right]\textrm{i}t}
\right).\]
At $p=q=-1$, $\mathcal{W}_1=\mathcal{W}_2=1$. Thus,
\begin{eqnarray} \label{1m0G}
\frac{1}{m^{(0)}}\left. \mathcal{G} m^{(0)}\right|_{p=q=-1}= \frac{(-2)}{p(e^{\kappa})+q(e^{\lambda})} \exp\left({\frac{1}{2}\left[p(e^{\kappa})+q(e^{\lambda})+2\right]x-\frac{1}{4}\left[p^2(e^{\kappa})-q^2(e^{\lambda})\right]\textrm{i}t}\right).
\end{eqnarray}
We need to expand the right side of this equation into double Taylor series in $\kappa$ and $\lambda$. To expand the fraction in front of the exponential term, we notice that for any functions $f(\kappa)$ and $g(\lambda)$,
\begin{eqnarray}
&&\frac{-2}{f(\kappa)+g(\lambda)}=\frac{-2\left[f(0)+g(0)\right]}{\left[f(\kappa)+g(0)\right]\left[g(\lambda)+f(0)\right]}
\frac{1}{1-\frac{f(\kappa)-f(0)}{f(\kappa)+g(0)}\hspace{0.05cm}
\frac{g(\lambda)-g(0)}{g(\lambda)+f(0)}} \nonumber \\
&& =\exp\left(-\ln \frac{\left[f(\kappa)+g(0)\right]\left[g(\lambda)+f(0)\right]}{-2\left[f(0)+g(0)\right]}\right)
\sum_{\nu=0}^{\infty}
\left[ \frac{f(\kappa)-f(0)}{f(\kappa)+g(0)}\hspace{0.05cm}\frac{g(\lambda)-g(0)}{g(\lambda)+f(0)} \right]^{\nu}.
\end{eqnarray}
Thus, substituting
\[
f(\kappa)=p(e^{\kappa})=c_1e^{\kappa/3}+c_2e^{-\kappa/3}, \quad g(\lambda)=q(e^{\lambda})=c_1e^{\lambda/3}+c_2e^{-\lambda/3}
\]
into the above equation, we get
\begin{eqnarray}
&&\exp\left(-\ln \frac{\left[f(\kappa)+g(0)\right]\left[g(\lambda)+f(0)\right]}{-2\left[f(0)+g(0)\right]}\right) \nonumber \\
&& = \exp\left(-\ln \left[\frac{1}{2}-\cosh\left(\frac{\kappa}{3}+ \frac{2\textrm{i}\pi}{3}\right)\right] -\ln \left[ \frac{1}{2}-\cosh\left(\frac{\lambda}{3}+ \frac{2\textrm{i}\pi}{3}\right)\right] \right) \nonumber \\
&& = \exp \left( \sum_{k=1}^{\infty} r_{k}\left(\kappa^k + \lambda^{k}\right)\right), \label{firstfraction1}
\end{eqnarray}
where $r_{k}$ are coefficients of the Taylor expansion given by (\ref{skrkexpcoeff1}) in Theorem 2. In addition, we get
\begin{eqnarray}
&& \sum_{\nu=0}^{\infty}
\left[ \frac{f(\kappa)-f(0)}{f(\kappa)+g(0)}\hspace{0.05cm}\frac{g(\lambda)-g(0)}{g(\lambda)+f(0)} \right]^{\nu}  \nonumber \\
&&=\sum_{\nu=0}^{\infty} \left( -\frac{\kappa\lambda}{12} \right)^{\nu}
\exp\left( \nu \ln\left[
\frac{2\textrm{i}\sqrt{3}}{\kappa} \tanh\frac{\kappa}{6}\tanh\left(\frac{\kappa}{6}+ \frac{2\textrm{i}\pi}{3}\right)\right]
+\nu\ln \left[\frac{2\textrm{i}\sqrt{3}}{\lambda} \tanh\frac{\lambda}{6}\tanh\left(\frac{\lambda}{6}+ \frac{2\textrm{i}\pi}{3}\right)\right] \right) \nonumber \\
&& = \sum_{\nu=0}^{\infty} \left( -\frac{\kappa\lambda}{12} \right)^{\nu} \exp\left( \nu  \sum_{k=1}^{\infty} s_{k}\left(\kappa^k + \lambda^{k}\right) \right)\label{secondfraction2},
\end{eqnarray}
where $s_{k}$ are defined by (\ref{skrkexpcoeff2}) in Theorem 2. Regarding the exponential term in Eq. (\ref{1m0G}), when functions $p(\cdot)$ and $q(\cdot)$ in Eq. (\ref{CubicEqRoot2}) are inserted, we get
\begin{eqnarray}
&& \exp\left({\frac{1}{2}\left[p(e^{\kappa})+q(e^{\lambda})+2\right]x-\frac{1}{4}\left[p^2(e^{\kappa})-q^2(e^{\lambda})\right]\textrm{i}t}\right)
\nonumber \\
&& = \exp \left[\frac{1}{2}\left( c_{1} e^{\kappa/3}+c_{2} e^{-\kappa/3} + c_{1}e^{\lambda/3}+c_{2} e^{-\lambda/3} +2\right)x-\frac{1}{4}\left( c_{2} e^{2\kappa/3}+c_{1} e^{-2\kappa/3} -c_{2}e^{2\lambda/3} - c_{1} e^{-2\lambda/3}\right)\textrm{i}t\right]  \nonumber \\
&& = \exp\left(\sum_{k=1}^{\infty} \frac{\kappa^k}{k!}  \frac{c_{1}+(-1)^k c_{2}}{2\cdot 3^k} \left[x +(-2)^{k-1} \textrm{i}t\right] +
     \sum_{k=1}^{\infty} \frac{\lambda^k}{k!} \frac{c_{1}+(-1)^k c_{2}}{2\cdot 3^k} \left[x -(-2)^{k-1} \textrm{i}t\right]\right).
\end{eqnarray}
Combining all these results, Eq. (\ref{1m0G}) reduces to
\begin{eqnarray*}
\frac{1}{m^{(0)}}\left. \mathcal{G} m^{(0)}\right|_{p=q=-1}
= \sum_{\nu=0}^{\infty} \left( -\frac{\kappa\lambda}{12} \right)^{\nu} \exp\left(  \sum_{k=1}^{\infty} \left(x_{k}^{+}+\nu s_{k}\right) \kappa^k + \sum_{k=1}^{\infty} \left(x_{k}^{-}+\nu s_{k}\right) \lambda^{k} \right),
\end{eqnarray*}
where $x_{k}^{\pm}$ are given by (\ref{skrkexpcoeff}) in Theorem 2. Then, taking the coefficients of $\kappa^k \lambda^l $ on both sides, we find
\[
\frac{1}{m^{(0)}} \frac{1}{k! l!}  \left[f_1(p)\partial_{p}\right]^{k}  \left[f_2(q)\partial_{q}\right]^{l} \left. m^{(0)}\right|_{p=-1, q=-1}
=\sum_{\nu=0}^{\min(k,l)}  \left( -\frac{\kappa\lambda}{12} \right)^{\nu} S_{k-\nu}\left( \textbf{\emph{x}}^{+}+\nu \textbf{\emph{s}} \right)
S_{l-\nu}\left( \textbf{\emph{x}}^{-}+\nu \textbf{\emph{s}} \right).
\]
From this result, we get
\[
\frac{1}{m^{(0)}}\mathcal{A}_{i} \mathcal{B}_{j} \left. m^{(0)}\right|_{p=-1, q=-1}=\sum_{\nu=0}^{\min(i,j)} \Phi_{i\nu} \Psi_{j\nu},
\]
where $\Phi_{i\nu}$ and $\Psi_{j\nu}$ are defined in Eqs. (\ref{phipsiexp1})-(\ref{phipsiexp2}) of Theorem 2. Since the Boussinesq solution (\ref{Th1-solutions}) is invariant when the function $\sigma$ is multiplied by an exponential factor of a linear function in $x$ and $t$, the above expression (with the exponential factor $1/m^{(0)}$) as the matrix element of $\sigma$, rather than the original $\mathcal{A}_{i} \mathcal{B}_{j} m^{(0)}$ in Eq. (\ref{mijtheorem1}), would give the same solution. This finishes the proof of Theorem 2.

In the end, we show the boundary conditions (\ref{BoundCondition}) of these rogue waves. For this purpose, we first rewrite the determinant (\ref{SigmanAlg}) of $\sigma(x,t)$ with Schur-polynomial matrix elements in Theorem 2 into a $3N \times 3N$ determinant, as was done for the NLS equation in \cite{OhtaJY2012}. Using this larger determinant, one can directly obtain the leading-order terms in the polynomial function $\sigma(x,t)$ as
\[
\sigma(x,t) = c_N \left( x^2+t^2 \right)^{N(N+1)/2} + \mbox{lower-degree terms},
\]
where $c_N$ is a $N$-dependent constant. From this result, one can see quickly that the solution $u(x,t)$ in Eq. (\ref{Th1-solutions}) satisfies the boundary conditions (\ref{BoundCondition}).

\section{Conclusion and Discussion}
In this article, we have derived rogue wave solutions in the Boussineq equation (\ref{BoussiEq}) through the bilinear KP-reduction method, and these solutions are given explicitly as Gram determinants with matrix elements in terms of Schur polynomials. Our solutions contain more free parameters than those reported before \cite{ClarksonDowie2017}, and they exhibit new interesting wave patterns. We have also shown that the rogue wave of maximum amplitude at each order is generally asymmetric in space, which is quite unusual in integrable equations.

Technically, our main contribution to the bilinear KP-reduction method for rogue waves is a new judicious choice of differential operators for matrix elements (the $\mathcal{W}$-$p$ treatment). Compared to the previous choices, this $\mathcal{W}$-$p$ treatment drastically simplifies the dimension reduction calculation as well as the analytical expressions of rogue wave solutions.

Can this $\mathcal{W}$-$p$ treatment be applied to other integrable equations, such as the NLS-Boussinesq equation and the Yajima-Oikawa system considered in \cite{XiaoeYong2018,Chen_Juntao}? The answer is a definite yes, maybe with minor modifications possibly. For instance, for the Yajima-Oikawa system, we can slightly modify our definitions of the $\mathcal{W}_1(p)$ and $\mathcal{W}_2(q)$ functions as
$Q_k=\beta_k [\mathcal{W}_k+\mathcal{W}_k^{-1}]$, where $Q_1(p)$ and $Q_2(q)$ are certain functions arising from the dimension reduction of the Yajima-Oikawa system, and $\beta_1$, $\beta_2$ are certain constants. For these $\mathcal{W}_1(p)$ and $\mathcal{W}_2(q)$ functions, the differential operators (\ref{003b}) with $f_1(p)$ and $f_2(q)$ defined in (\ref{f1pf2q}) will again simplify the dimension reduction calculation and produce rogue wave expressions similar to those in this article. Comparatively, the rogue wave expressions derived in \cite{XiaoeYong2018,Chen_Juntao} with old choices of differential operators had to involve additional indices in the coefficients of the differential operators and matrix elements, and those coefficients had to be obtained by complicated recurrence relations. This $\mathcal{W}$-$p$ treatment can also be applied to the NLS equation. In this special case, we would get $f_1(p)=p$ and $f_2(q)=q$, which reproduces the old differential operators used in Ref. \cite{OhtaJY2012}. Thus, this $\mathcal{W}$-$p$ treatment is a general and useful technique to streamline the bilinear derivation of rogue waves in integrable systems when the dimension reduction is needed.


\section*{Acknowledgement}

This material is based upon work supported by the Air Force Office of Scientific
Research under award number FA9550-18-1-0098 and the National Science Foundation under award number DMS-1616122.

\end{document}